\UseRawInputEncoding
\documentclass{article}
\usepackage{amsmath}
\usepackage{amsthm}
\usepackage{amsfonts}
\usepackage{graphicx}
\usepackage{lscape}
\usepackage{rotating}
\usepackage{longtable}
\usepackage{extarrows}
\usepackage{float}
\usepackage{hyperref}
\usepackage{geometry}
\usepackage{authblk}
\usepackage{verbatim}
\usepackage{mathtools}
\usepackage{tikz-feynman}
\usepackage{tikz}
\usetikzlibrary{shapes.misc}
\usetikzlibrary{decorations.pathmorphing}
\tikzset{snake it/.style={decorate, decoration=snake}}

\title{Scrambling with conservation laws}
\author[1]{Gong Cheng}
\affil[1]{Condensed Matter Theory Center and Department of Physics, University of Maryland, College Park, MD 20742, USA}
\author[1,2]{Brian Swingle}
\affil[2]{Department of Physics, Brandeis University, Waltham, MA 02453, USA}

\date{}

\begin{document}
\maketitle

\begin{abstract}
In this article we discuss the impact of conservation laws, specifically $U(1)$ charge conservation and energy conservation, on scrambling dynamics, especially on the approach to the late time fully scrambled state. As a model, we consider a $d+1$ dimensional ($d\geq 2$) holographic conformal field theory with Einstein gravity dual. Using the holographic dictionary, we calculate out-of-time-order-correlators (OTOCs) that involve the conserved $U(1)$ current operator or energy-momentum tensor. We show that these OTOCs approach their late time value as a power law in time, with a universal exponent $\frac{d}{2}$. We also generalize the result to compute OTOCs between general operators which have overlap with the conserved charges.
\end{abstract}

\tableofcontents

\section{Introduction}
\subsection{Motivation}
Quantum information scrambling~\cite{Hayden:2007cs,Sekino2008,Brown_2012_Scrambling} is a fundamental phenomenon in chaotic many-body systems that has been under wide discussion in recent years. On the theoretical side, this activity focused on the study of out-of-time-order correlators (OTOCs)~\cite{Larkin1969,Shenker2014,Shenker2014a,kitaev2015,Aleiner2016,Hosur2016,Hallam:2019npn,Gu2017,Luitz2017,Patel2017,Nahum2017a,VonKeyserlingk2017,Xu2018a}. For a chaotic system with large $N$ number of degrees of freedom per unit volume, the OTOC displays an exponentially increasing deviation from its initial value which is characterized by a quantum Lyapunov exponent. This period of growth occurs after local equilibrium is achieved but before the scrambling time, when the system approaches global equilibrium. Schematically, given simple Hermitian operators $W$ and $V$, one has

\begin{equation}\label{ineq1}
\text{OTOC} = \langle W(t) V(0) W(t) V(0)\rangle \sim f_0-\frac{f_1}{N}e^{\lambda_L t}+\cdots.
\end{equation}

For a large $N$ conformal field theory (CFT) holographically described by Einstein gravity, earlier works \cite{Shenker2014}\cite{Shenker2014a}\cite{2dcft}\cite{Shenker2014b}\cite{Roberts2015} calculated OTOCs through geometric methods, by studying shockwave geometries. The Lyapunov exponent in such a theory is equal to $\frac{2\pi}{\beta}$, saturating the conjectured chaos bound \cite{bound}. For times much larger than scrambling time (the late time regime), the OTOC decays to zero exponentially fast, with a different but related exponent. By contrast, it was shown in \cite{conservation} that in a random circuit model with local charge conservation law, the OTOC between the charge density operator and a non-conserved operator displays a power law tail at late time. While such power law tails are expected to be generic, they have not yet been seen in holographic systems. This missing piece of physics motivated the present study. Other studies of the interplay between conservation laws, hydrodynamics, and OTOCs include \cite{Lucas:2017ibu,Blake_2016,Blake_2017,Bohrdt_2017,Grozdanov_2018,Blake_2018,Blake_2018_1,Rakovszky_2018,Haehl_2018,Marino_2019,Grozdanov_2019,Chen_2020,ramirez2020chaos,Abbasi_2020}.

The bulk of this work focuses on the $U(1)$ case: we compute OTOCs between the conserved charge density and a (non-conserved) scalar operator and show that holographic systems also exhibit power law tails consistent with the random circuit result. We also argue that power law tails will be induced by energy conservation. This has not been shown before, but it is important since Hamiltonian systems generically have energy conservation but not charge conservation. In the remainder of the introduction, we review existing holographic calculations, focusing on the scattering approach. Then, in the rest of the article, we show how these calculations are modified due to $U(1)$ charge conservation and we interpret the result physically. We also discuss the extension to energy conservation. Overall, we view this work as a study of the interplay between slow modes, as in hydrodynamics, and the fast dynamics of scrambling.

\subsection{Review of the holographic calculation}
The holographic calculation of OTOCs for scalar operators is discussed in many works. The approach that we follow here is based on the scattering approach discussed in~\cite{Shenker2014a}.  In this approach, the boundary OTOC is written as an inner product of $in$ and $out$ asymptotic states.
\begin{equation}
\begin{split}
    \langle W(t_1,x_1)&V(t_2,x_2)W(t_1,x_1)V(t_2,x_2)\rangle=\langle out|in \rangle\\
    &|in\rangle=W(t_1,x_1)V(t_2,x_2)|TFD\rangle\\
    &|out\rangle =V^{\dagger}(t_2,x_2)W^{\dagger}(t_1,x_1)|TFD\rangle.
\end{split}
\end{equation}

From the bulk perspective, these $in$ and $out$ states may be written in terms of bulk wavefunctions as
\begin{equation}\label{ineq3}
\begin{split}
    |in\rangle=\int dp^u dx\int dp^v dx' \phi_W(p^v,x)\phi_V(p^u,x')|p^u,x,p^v,x'\rangle_{in}\\
    |out\rangle=\int dp^udx\int dp^vdx' \phi_{W^{\dagger}}(p^v,x)\phi_{V^{\dagger}}(p^u,x')|p^u,x,p^v,x'\rangle_{out}.
\end{split}
\end{equation}
Here $u$ and $v$ are null coordinates in the black hole geometry dual to the thermofield double state $|TFD \rangle$.

\begin{figure}
\center
\begin{tikzpicture}
\draw (0,-3)--(0,3);
\draw (4,-3)--(4,3);
\draw[snake it] (0,3)--(4,3);
\draw[snake it] (0,-3)--(4,-3);
\draw (0,3)--(4,-3);
\draw (0,-3)--(4,3);
\draw[->] (3.9,-2.5)node[right]{$V(t_2)$} --(3.65,-2.125);
\draw   (3.65,-2.125)--(3.4,-1.75);
\draw[->] (0.56,-2.5)--(0.81,-2.125);
\draw   (0.81,-2.125)--(1.06,-1.75);
\draw   (0.56,2.5)--(0.81,2.125);
\draw[->] (1.06,1.75)--(0.81,2.125);
\draw   (3.9,2.5)node[right]{$W(t_1)$}--(3.65,2.125);
\draw[->]  (3.4,1.75)--(3.65,2.125);
\draw[gray] plot [smooth] coordinates {(0,-2.3) (1,-2.2) (2,-2.3) (3,-2.4) (4,-2.3)};
\draw[gray] plot [smooth] coordinates {(0,2.3) (1,2.2) (2,2.3) (3,2.4) (4,2.3)};
\draw (2,-2.3) node[below] {$|in>$};
\draw (2,2.3) node[above] {$|out>$};
\end{tikzpicture}
\caption{$in$ and $out$ sates in the Penrose diagram. The data on the $in$ and $out$ slices are given by Eq.~\eqref{ineq3}.}
\label{infig1}
\end{figure}
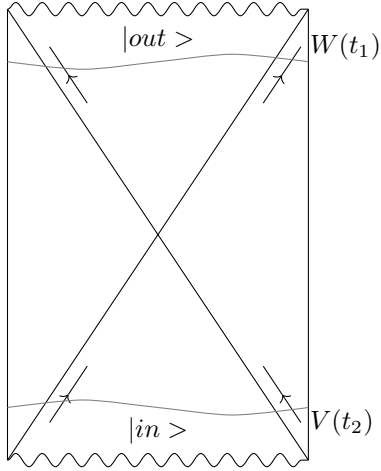

Note that the momentum states are only well-defined near the black hole horizon. We think of the scattering as taking place in the approximately flat near horizon region as described by Kruskal coordinates. This corresponds to the case of large time separation between $t_1$ and $t_2$, as shown in Fig~\ref{infig1}. The wave functions are given by bulk-to-boundary propagators. So
\begin{equation}
\begin{split}
    &\phi_W(p^v_1,x)=\int dv e^{ip_1^u v}\langle \phi_W(u,v,x)W(t_2,x_2)\rangle|_{u=-\epsilon}\\
    &\phi_V(p^u_2,x')=\int du e^{ip^v_2 u}\langle \phi_V(u,v,x)V(t_1,x_1)\rangle|_{v=\epsilon}\\
    &\phi_{W^{\dagger}}(p^v_1,x)=\int dv e^{ip^u_1 v}\langle \phi_W(u,v,x)W^{\dagger}(t_2,x_2)\rangle|_{u=-\epsilon}\\
    &\phi_{V^{\dagger}}(p^u_2,x')=\int du e^{ip^v_2 u}\langle\phi_V(u,v,x)V^{\dagger}(t_1,x_1)\rangle|_{v=\epsilon}\\
\end{split}
\end{equation}

These formulae have a direct interpretation as bulk scattering states sourced by boundary operators. The relevant energy scale is determined by the boundary time separation through the Mandelstam variable $s:=2p_1^vp_2^u\sim e^{\frac{2\pi}{\beta}t_{12}}$. For scrambling physics, we are interested in time scales that are larger than the relaxation time. The dominate contribution in this regime comes from summing ladder diagrams with graviton exchanges \cite{Eikonal}. To leading order in $s$, the S-matrix approaches a pure phase, obtained from the eikonal approximation,
\begin{equation}
|p_1^u,x,p_2^v,x'\rangle_{out}\sim e^{i\delta(s,b)}|p_1^u,x,p_2^v,x'\rangle_{in}+\cdots
\end{equation}

The physical interpretation of this phase factor is interaction of particle $1$ with a gravitational shockwave sourced by particle $2$~\cite{Dray:1985yt}\cite{Dray:1984ha}. The shockwave metric is
\begin{equation}
    ds^2=A(uv)du[dv-\delta(u)h(x)du]+B(uv)d^dx.
\end{equation}
While passing through the shockwave located near $u\sim 0$, the scalar wave function receives a jump in the $v$ coordinate,
\begin{equation}
    \int dp_1^u\phi(p_1^u,x)e^{ip_1^uh(x)}e^{-ip_1^u v}=\phi(v-h(x),x).
\end{equation}
The phase $\delta(s,b)$ is then identified with the displacement factor $h(x) =G_N p_2^v \frac{e^{-\mu |x|} }{|x|^{\frac{d-1}{2}}}$.

For scalar operators, the OTOC is evaluated in earlier works. In the limit $\Delta_{W}\gg \Delta_V \gg 1$, the heavy particle's wavefunction is not much affected by the shockwave sourced by the light particle. So the amplitude is simply an inner product between $V$ particle wavefunctions with and without the shockwave,
\begin{equation}\label{ineq2}
\begin{split}
   \text{OTOC}&\sim \int dvdx \phi_V(v,x)\partial_v\phi_V(v-h(x),x)\\
   &\sim \left[\frac{1}{1+\frac{G_N\Delta_W}{\Gamma} e^{\frac{2\pi}{\beta}t-\mu |x|}}\right]^{\Delta_V}.
\end{split}
\end{equation}
$\Gamma$ is a constant depending on the regularization of the correlator.

At early time, when the second term in denominator is much smaller than the first term, the OTOC is decreasing as shown in Eq.~\eqref{ineq1}. Note that $e^{\frac{2\pi}{\beta}t}$ is roughly the colliding energy, and it enters through $p$ dependence of $h(x)$. The exponent is $\frac{2\pi}{\beta}$, independent of the operator. The correlator has decayed significantly when the second term in the denominator is $O(1)$. After this time, the correlator experiences an expontial decay with an operator-dependent exponent. We refer to this as the late time regime of OTOC.

\subsection{Results and outline}
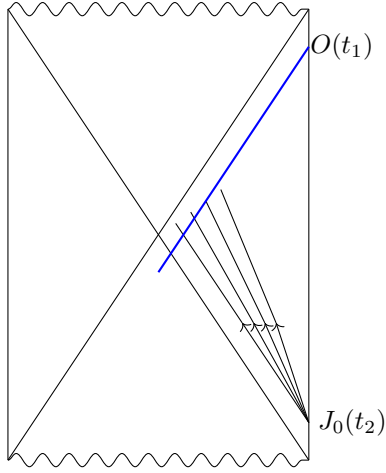
\begin{figure}
\center
\begin{tikzpicture}
\draw (0,-3)--(0,3);
\draw (4,-3)--(4,3);
\draw[snake it] (0,3)--(4,3);
\draw[snake it] (0,-3)--(4,-3);
\draw (0,3)--(4,-3);
\draw (0,-3)--(4,3);
\draw[->] (4,-2.5)node[right]{$J_0(t_2)$} --(3.115,-1.175);
\draw   (3.115,-1.175)--(2.23,0.15);
\draw[->] (4,-2.5)--(3.265,-1.175);
\draw  (3.265,-1.175)--(2.43,0.3);
\draw[->] (4,-2.5)--(3.415,-1.175);
\draw  (3.415,-1.175)--(2.63,0.45);
\draw[->] (4,-2.5)--(3.565,-1.175);
\draw   (3.565,-1.175)--(2.83,0.6);
\draw   (3.9,2.5)node[right]{$O(t_1)$};
\draw[blue, thick] (2,-0.5)--(4,2.5);
\end{tikzpicture}
\caption{Photon scatters with shockwave. The blue line denotes the shockwave sourced by scalar operator. The photon sourced by charge operator spreads in space-time. Classically, we can view it as a bunch of photons, with different longitudinal energy.   }
\label{infig2}
\end{figure}

The situation is changed when we consider an OTOC that involves conserved charges. For example, the R-charge in $\mathcal{N}=4$ SYM theory, and more generally the energy-momentum tensor. We will see that due to the hydrodynamical property of the conserved current, the particle sourced by these operators in the bulk spreads over a large region of space-time. As a result, the collision responsible for scrambling happens at a wide range of space-time points in the classical picture (see Fig~\ref{infig2}). When the collision occurs near the horizon, the center of mass energy is large, but when the collision occurs further away from the horizon, the center of mass energy is smaller. This leads to a smearing of the exponential factor in Eq.~\eqref{ineq2}, effectively replacing the OTOC formula with
\begin{equation}
  \text{OTOC}\sim  \int_0^{+\infty} ds \frac{1}{s^{\frac{d}{2}+1}}\left[\frac{1}{1+\frac{c}{N}e^{\frac{2\pi}{\beta}(t-s)-\mu|x|}}\right]^{\alpha}
\end{equation}
where $c$ and $\alpha$ are some constants. One can then show that at late time, the OTOC becomes $\sim t^{-\frac{d}{2}}$.

The article is organized as follows. For simplicity, we start with the conserved charge density of a $U(1)$ symmetry. In section~\ref{sec2} we give an expression for the dual photon's wave function, to lowest order in transverse momentum and frequency. Then we analyze the inner product and equation of motion for the photon field. Using these ingredients and some additional approximations, we calculate the OTOC. In section~\ref{sec3} we interpret these calculations in terms of the  classical picture in Fig.~\ref{infig2}. In section~\ref{sec4} we  discuss the case of the energy-momentum tensor. Finally, we generalize the result in section~\ref{sec5} to the case of generic operators with overlap with the conserved currents.

\section{Correlator with U(1) charge}\label{sec2}
\subsection{Solution to equation of motion }
In an AdS-Schwarzschild black hole background, the metric is
\begin{equation}\label{eq0}
    ds^2=L^2\left[-\frac{f(R)}{R^2}dt^2+\frac{1}{R^2 f(R)}dR^2+\frac{1}{R^2} d^d \vec{x}\right]
\end{equation}

where $R=R_+$ is the horizon and boundary is at $R=0$. The inverse temperature is $\beta=\frac{4\pi R_+}{d+1}$ and $f(R)=1-(\frac{R}{R_+})^{d+1}$. It's more convenient to use the tortoise coordinate, $r$, defined by
\begin{equation}\label{eq01}
    dr=-\frac{1}{f(R)}dR.
\end{equation}
The domain of $r$ is $r\in (-\infty,0 ]$ with $-\infty$ corresponding to the horizon and $0$ corresponding to the boundary. In this coordinate, the metric components are  $g_{rr}=-g_{tt}=L^2\frac{f(R)}{R^2}$. Here we use $\vec{x}$ to denote the coordinates of the transverse directions. We also consider a Maxwell field propagating in this geometry whose dynamics determines the behavior of a $U(1)$ current operator $J_0$ on the boundary. In this note, we will focus on the four-point correlation function with two insertions of $J_0$ and two insertions of a scalar operator $O$.

The Maxwell-Einstein equation is
\begin{equation}\label{eqmaxein}
    \partial_{\mu}(\sqrt{-g}F^{\mu\nu})=\partial_{\mu}(\sqrt{-g}g^{\mu\rho}g^{\nu\sigma}F_{\rho\sigma})=0.
\end{equation}
Consider the ansatz $A_{i} =0$ for all the transverse directions. The Fourier mode decomposition is
\begin{equation}
    A_{\mu}(r,t,\vec{x})=\int d\omega d\vec{k} A_{\mu}(r,\omega,k)e^{-i\omega t}e^{i \vec{k}\cdot \vec{x}}
\end{equation}
Using the trick in \cite{membrane}, we pick a transverse coordinate frame for each $\vec{k}$, such that the $x$-axis is parallel to $\vec{k}$ direction, with the other axes perpendicular to it. Then $\partial_x$ can be replaced by $ik$ when acting on that mode, and derivatives in other transverse direction are replaced by $0$. We also assume that the $A$ field is spherically symmetric and excited by a point source sitting at $\vec{x'}=0$, so the momentum space $A$ only depends on norm of $\vec{k}$.

Written in component form, the equations are
\begin{equation}
   \Bigg\{
   \begin{array}{ccc}
    \partial_r(\sqrt{-g}g^{rr}g^{tt}(\partial_r A_t-\partial_t A_r))+\partial_x(\sqrt{-g}g^{xx}g^{tt}\partial_x A_t)=0 \\
   \partial_t(\sqrt{-g}g^{tt}g^{rr}(\partial_t A_r-\partial_r A_t))+\partial_x(\sqrt{-g}g^{xx}g^{rr}\partial_x A_r)=0 \\
   \partial_r(\sqrt{-g}g^{rr}g^{xx}(-\partial_x A_r))+\partial_t(\sqrt{-g}g^{tt}g^{xx}(-\partial_x A_t))=0.
   \end{array}
\end{equation}
From the second equation, one can deduce
\begin{equation}\label{eq1}
\begin{split}
    -\omega^2 g^{tt} A_r+i\omega g^{tt}\partial_r A_t-k^2 g^{xx}A_r & =0\\
    \implies  A_r=\frac{i\omega g^{tt}\partial_r A_t}{\omega^2g^{tt}+k^2g^{xx}}
\end{split}
\end{equation}
Then one can use the first equation to find a second order differential equation for $A_t$,
\begin{equation}
     \partial_r(\sqrt{-g}g^{rr}g^{tt}\frac{k^2g^{xx}\partial_r A_t}{\omega^2g^{tt}+k^2g^{xx}})-k^2\sqrt{-g}g^{xx}g^{tt}A_t=0.
\end{equation}

This equation has two independent solutions. As $r\rightarrow \infty$, they behave like $e^{\pm i\omega r}$, corresponding to out-going and in-falling boundary condition at the horizon. We focus on one of them, since the other one is obtained by complex conjugation. The solution can be found explicitly in the small frequency and small momentum limit by setting $\omega\rightarrow \lambda \omega$, $k\rightarrow \lambda k$, and taking $\lambda \ll 1$. Up to first order in $\lambda$, we have
\begin{equation}
    A_t(r,\omega,k)= e^{-i \omega r}C(\omega,k)[1+i\omega \int_{-\infty}^r dr(1- (\frac{R}{R_+})^{d-2})+\frac{ik^2}{\omega} \int_{-\infty}^r dr (\frac{R}{R_+})^{d-2}f+ O(\lambda^2)].
\end{equation}
The normalization constant $C$ is fixed by requiring $\lim_{r\rightarrow 0}A_t(r,\omega,k)\rightarrow 1$. Then we find that
\begin{equation}\label{eq(9)}
    A_t(r,\omega,k)=e^{-i\omega r}\left[\frac{1+i\omega H(r)+i\frac{k^2}{\omega}R_+\frac{1-(\frac{R}{R_+})^{d-1}}{d-1}}{1+i\omega H(0)+i\frac{k^2}{\omega}R_+\frac{1}{d-1}}+O(\lambda^2)\right]
\end{equation}

The factor $\frac{1}{d-1}R_+$ is identified as the diffusion constant $D$. $H(r)$ is the indeterminate integral of the second term in Eq.~\eqref{eq(9)}. If we ignore the $\omega H(0)$ term in the denominator, then this function has a pole  at $\omega = -iD k^2$. With this $\omega \sim k^2$ scaling, the $\omega H(0) $ is indeed subleading compared to the $i k^2 D/\omega$ term at small $\omega$ and $k$. Moreover, we can anticipate that $\omega^2$ is of order $k^4$ after using the residue theorem.  Hence, to the leading order in small $k$ and $\omega$, we can neglect $\omega^2$ and higher order terms. Another approximation is to set $(\frac{R}{R_+})\sim 1$, corresponding to the near horizon region. Since the scrambling time is large, the relevant physics indeed happens within this region.  In summary, we have

\begin{equation}\label{eq2}
    A_t(r,\omega,k)\sim e^{-i\omega r}\frac{\omega}{\omega+iDk^2}.
\end{equation}

The real space form is
\begin{equation}\label{eq10}
    A_t(r,t, \vec{x})\sim \partial_t \left[\frac{1}{(t-t'+r)^{\frac{d}{2}}}e^{-\frac{|\vec{x}-\vec{x}'|^2}{4D(t-t'+r)}}\theta(t-t'+r)\right]
\end{equation}
where $t'$ and $\vec{x}'$ label the position of the boundary source. We expect this expression to hold near the horizon when the transverse separation from the source is large. Now we use Eq.~\eqref{eq1} to obtain the other components,
\begin{equation}\label{eq3}
\begin{split}
    A_r &\sim \frac{i}{\omega}\partial_r A_t =A_t,\\
    F_{tr}=\partial_t A_r -\partial_r A_t&=\frac{-k^2 g^{xx}\partial_r A_t}{\omega^2 g^{tt}+k^2 g^{xx}}\sim \frac{i k^2}{\omega}g^{xx}g_{tt}A_t.
\end{split}
\end{equation}
Since $g_{tt} \rightarrow 0$ near horizon, $F_{tr}$ is also very small there. (Note that the physical electric field $E_r= \sqrt{g^{tt}}\sqrt{g^{rr}}F_{tr}$ is still finite).

\subsection{Inner product of gauge field}
The gauge invariant inner product between two gauge field configurations $A_1$ and $A_2$ is

\begin{equation}\label{eq12}
    ( A_1,A_2)=\int \sqrt{h} n^{\mu} ({F_1}^*_{\mu\nu} {A_2}^{\nu}-{F_2}_{\mu\nu}{A_1^*}^{\nu})
\end{equation}
Where the integration is over a Cauchy surface.  Following \cite{Shenker2014a}, we choose to integrate on constant $u\sim 0$ slice. Then this inner product becomes
\begin{equation}\label{eq11}
    \int dvd\vec{x} \left(\frac{R}{L}\right)^{-d} g^{uv} ({F_1}_{vu}^* {A_2}_{v}- {F_2}_{vu}{A_1^*}_{v} ),
\end{equation}
with $u$ and $v$ related to $t$ and $r$ by
\begin{equation}
\begin{split}
    u=-e^{\frac{2\pi}{\beta}(r-t)},\\
    v=e^{\frac{2\pi}{\beta}(t+r)},\\
    (\frac{\beta}{2\pi})^2g^{uv}=2uvg^{tt}.
\end{split}
\end{equation}
The various components of gauge field are related to the old ones by
\begin{equation}
\begin{split}
    A_v=&\frac{\partial t}{\partial v} A_t+ \frac{\partial r}{\partial v}A_r,\\
    =&\frac{\beta}{2\pi}\frac{1}{2v}(A_t+A_r),\\
     F_{uv}=&\frac{\partial t}{\partial u}\frac{\partial r}{\partial v}F_{tr}+\frac{\partial r}{\partial u}\frac{\partial t}{\partial v}F_{rt},\\
    =&-(\frac{\beta}{2\pi})^2\frac{1}{2uv}F_{tr}.
\end{split}
\end{equation}

Following Eq.~\eqref{eq2} and Eq.~\eqref{eq3},
\begin{equation}\label{eq4}
\begin{split}
    A_v=\frac{\beta}{2\pi}\frac{1}{v}A_t=&\frac{1}{v} v\partial_v \left(\int d\omega d\vec{k} \frac{i}{\omega+iDk^2}v^{-i\frac{\beta\omega}{2\pi}}e^{i \vec{k}\cdot \vec{x}}\right)\\
    \coloneqq & \partial_v \phi(v,\vec{x}),\\
    g_{xx}g^{tt}F_{tr}=&\int d\omega d\vec{k} \frac{i k^2}{\omega+iD k^2}v^{-i\frac{\beta\omega}{2\pi}}e^{i\vec{k}\cdot \vec{x}}=-\vec{\nabla}^2 \phi(v,\vec{x}).
\end{split}
\end{equation}
Plug this into the inner product expression Eq.~\eqref{eq11}, we get
\begin{equation}\label{eq6}
\begin{split}
    (A_1,A_2)=&\int dv d^d\vec{x} \left(\frac{R}{L}\right)^{2-d} [-\nabla^2\phi_1^*(v,\vec{x})\partial_v \phi_2(v,\vec{x})+\nabla^2\phi_2(v,\vec{x})\partial_v \phi_1^*(v,\vec{x})]\\
    =&\int dv d^d\vec{x} \left(\frac{R}{L}\right)^{2-d} [\vec{\nabla}\phi_1^*(v,\vec{x})\cdot\partial_v \vec{\nabla}\phi_2(v,\vec{x})-\vec{\nabla}\phi_2(v,\vec{x})\cdot\partial_v \vec{\nabla}\phi_1^*(v,\vec{x})]
\end{split}
\end{equation}
where $\phi(v,\vec{x})$ is the diffusion kernel given in Eq.~\eqref{eq4}. We will see that the inner product written in this way is very convenient when discussing the shockwave.

\subsection{Scattering states}

The original definition of OTOC is complicated by a UV divergence arising from coincident operator insertions. To avoid this, one can consider a regularized version obtained by inserting operators at different values of imaginary time,
\begin{equation}
    \mathcal{C}_{reg}=Tr[\rho^{1-\epsilon}V(t_2,\vec{x}_2)W(t_1,\vec{x}_1)\rho^{\epsilon}V(t_2,\vec{x}_2)W(t_1,\vec{x}_1)]
\end{equation}
To simplify the discussion, we choose a symmetric regularization. In the following, we will consider the correlator
\begin{equation}\label{eqhydrootoc}
    \mathcal{C_{J_0 O}}=Tr[\rho^{\frac{1}{2}}J_0(t_2,\vec{x}_2)O(t_1,\vec{x}_1)\rho^{\frac{1}{2}}J_0(t_2,\vec{x}_2)O(t_1,\vec{x}_1)].
\end{equation}
It has a representation as an inner product of $in$ and $out$ states
\begin{equation}\label{eq25}
\begin{split}
    |in\rangle=O^L(t_1,\vec{x}_1)J_0^{R}(t_2,\vec{x}_2)|S-AdS\rangle\\
    |out\rangle=J_0^L(t_2,\vec{x}_2)O^R(t_1,\vec{x_1})|S-AdS\rangle
\end{split}
\end{equation}
where $|S-AdS\rangle$ denotes the Schwarzschild-AdS black hole thermal double state,
\begin{equation}
    |S-AdS\rangle=\sum_n e^{-\frac{\beta}{2}E_n }|E_n\rangle_L|E_n\rangle_R.
\end{equation}

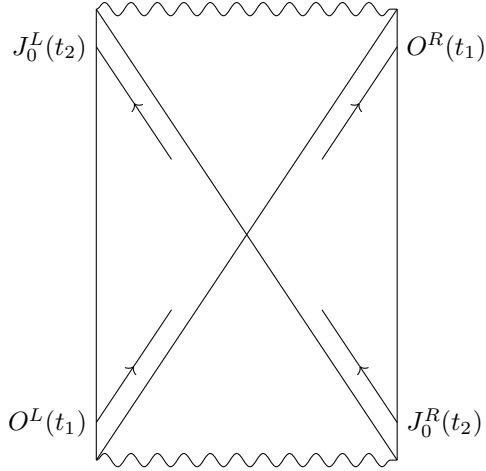
\begin{figure}[H]
\center
\begin{tikzpicture}
\draw (0,-3)--(0,3);
\draw (4,-3)--(4,3);
\draw[snake it] (0,3)--(4,3);
\draw[snake it] (0,-3)--(4,-3);
\draw (0,3)--(4,-3);
\draw (0,-3)--(4,3);
\draw[->] (4,-2.5)node[right]{$J_0^R(t_2)$} --(3.5,-1.75);
\draw   (3.5,-1.75)--(3,-1);
\draw[->] (0,-2.5)node[left]{$O^L(t_1)$}--(0.5,-1.75);
\draw   (0.5,-1.75)--(1,-1);
\draw   (0,2.5)node[left]{$J_0^L(t_2)$}--(0.5,1.75);
\draw[->] (1,1)--(0.5,1.75);
\draw   (4,2.5)node[right]{$O^R(t_1)$}--(3.5,1.75);
\draw[->]  (3,1)--(3.5,1.75);
\end{tikzpicture}
\caption{$in$ and $out$ sates in a symmetric regularization scheme}
\label{rung}
\end{figure}

The superscript $L$ and $R$ on each operator means that the excitation is created on the left or right side, respectively. For example, $J^R_0(t_2)$ creates a photon on the right. We denote the corresponding gauge field by $A_{\mu}^R$. In general it is a linear superposition of in-falling and out-going solutions, depending on the state we are constructing. Following  \cite{realtime}\cite{Son:2002sd}, we choose the coefficients such that the field has positive Kruskal frequency for in-falling mode and negative Kruskal frequency for out-going mode. For this purpose, we use the following combination
\begin{equation}
A_{\mu}^R(r,\omega,k)=(1+n(\omega)){A_{\mu}^R}_{\tiny{\text{in-falling}}}(r,\omega,k)-n(\omega){A_{\mu}^R}_{\text{in-falling}}^*(r,\omega,k),
\end{equation}
$n(\omega)$ is the Boltzmann factor.

For the in-falling part, we use the ansatz:
\begin{equation}
    \begin{split}
       {A^R_v}_{\text{in-falling}}&=\partial_v \phi^R_{\text{in-falling}}(v,\vec{x}),\\
        \phi^R_{\text{in-falling}}(v,\vec{x})&=\int \frac{d^dk}{(2\pi)^d} \frac{1}{(ve^{-\frac{2\pi}{\beta}t_2}-1)^{\frac{\beta Dk^2}{2\pi}}}  \theta(v-e^{\frac{2\pi}{\beta}t_2})e^{i\vec{k}\cdot  (\vec{x}-\vec{x_2})}.
    \end{split}
\end{equation}
This is just rewritten from Eq.~\eqref{eq4}, except that we have inserted an extra $1$ in the denominator. This change doesn't modify the long time behavior of the wave function, but it does provide some convenience in the analysis. On the other hand, the out-going part in $A_v$ is proportional to $u$. To evaluate the OTOC, we choose to calculate the inner product on the surface $u\sim 0$. Hence, the out-going mode's contribution can be neglected. The thermal factor $(1+n(\omega))$ is proportional to $\frac{1}{2\sin(\frac{\beta Dk^2}{2})}$, evaluated at the diffusive pole\footnote{The actual pole contains a small real part that is higher order in $k$ which keeps the integrand finite.}. It suggests the following ansatz for $A_v^R$
\begin{equation}
\begin{split}
A_v^R&=\partial_v \phi^R(v,\vec{x}),\\
    \phi^R(v,\vec{x})&=\int \frac{d^dk}{(2\pi)^d} \frac{1}{2\sin(\frac{\beta Dk^2}{2})}\frac{1}{(1-ve^{-\frac{2\pi}{\beta}t_2} )^{\frac{\beta Dk^2}{2\pi}}}  e^{i\vec{k}\cdot  (\vec{x}-\vec{x_2})}.
\end{split}
\end{equation}
A nice property of the above expression is that the real space form of $\phi^R$ and $\phi^R_{in-falling}$ are related by analytical continuation, if we treat it as a complete solution of equation of motion. To see this, note that
\begin{equation}
\frac{1}{(1-e^{\frac{2\pi}{\beta}(t-t_2+r+i0^+)} )^{\frac{\beta Dk^2}{2\pi}}}-\frac{1}{(1-e^{\frac{2\pi}{\beta}(t-t_2+r-i0^+)} )^{\frac{\beta Dk^2}{2\pi}}}=2i \sin(\frac{\beta Dk^2}{2}) \frac{1}{(ve^{-\frac{2\pi}{\beta}t_2}-1)^{\frac{\beta Dk^2}{2\pi}}}  \theta(v-e^{\frac{2\pi}{\beta}t_2})
\end{equation}

Similarly, on the left side boundary, the operator $J_0^L$ sources a wave function $A_v^L$. By symmetry, $A_v^L$ is related to $A_v^R$ via the transformation $(u,v)\rightarrow (-u,-v)$.  Correspondingly, we define $\phi^L(v,\vec{x})=\phi^R(-v,\vec{x})$. Then $A_v^L$ can be written as
\begin{equation}
\begin{split}
A_v^L&=\partial_v \phi^L(v,\vec{x}),\\
    \phi^L(v,\vec{x})&=\int \frac{d^dk}{(2\pi)^d} \frac{1}{2\sin(\frac{\beta Dk^2}{2})}\frac{1}{(1+ve^{-\frac{2\pi}{\beta}t_2} )^{\frac{\beta Dk^2}{2\pi}}}  e^{i\vec{k}\cdot (\vec{x}-\vec{x_2})}.
\end{split}
\end{equation}

The scalar operator $O^R$ and $O^L$ create scalar mode in the bulk. For simplicity, we assume that the scalar operator $O$ has a large conformal dimension, corresponding to a particle in bulk with a large mass. As a consequence, we can treat the scalar particle semi-classically. As it moves deep into the bulk, the scalar mode carries the shockwave along with it, which modifies the photon's wave function as we will analyze in the next section. In contrast, we will neglect the back-reaction from the photon on this scalar mode.

\subsection{Shockwave geometry}
Since we are interested in the large time limit of the OTOC, we will take $t_1 \gg \beta$ and $t_2 \ll -\beta$. As a result, the geodesic of the scalar particle is approximated by $u=\epsilon \sim0$. The shockwave geometry is described by a metric that contains a singularity near the horizon,
\begin{equation}
    ds^2=2g_{uv}du[dv-\delta(u)h(\vec{x})du]+g_{xx}d\Omega_{d},
\end{equation}
where $h(\vec{x})=\frac{\Delta_O}{N} e^{\frac{2\pi}{\beta}t_1-\mu |\vec{x}-\vec{x}_1|}$ and $\mu=\sqrt{\frac{2d}{d+1}}\frac{2\pi}{\beta}$.

The Maxwell-Einstein equation in this background is
\begin{equation}\label{eq5}.
   \begin{split}
       \partial_v(\sqrt{\gamma} F_{uv}g^{uv})+\sum_{i=1}^d \partial_i(\sqrt{\gamma}g^{ii}F_{iv})=0,\\
       \partial_u(\sqrt{\gamma}F_{vi}g^{ii})+\partial_v(\sqrt{\gamma}F_{ui}g^{ii})+\partial_{v}(\sqrt{-g}g^{vv}F_{vi}g^{ii})=0,
   \end{split}
\end{equation}
where the $i$'s run from $1$ to $d$ and label the transverse directions. The effect of the shockwave is encoded in the $g^{vv}$ component. $\gamma$ is the metric determinant in the transverse directions, so $-g=g_{uv}^2\gamma$.

As before, we simplify the equation by approximating $\frac{R}{R_+} \sim 1$ such that we can set $\gamma=(\frac{L}{R})^{2d}$ and $g^{ii}=(\frac{L}{R})^2$. Then using $g^{vv}=-(g^{uv})^2 g_{uu}=2g^{uv}h(\vec{x})\delta(u)$, the second equation becomes
\begin{equation}\label{eq7}
    -\partial_u\partial_iA_v-\partial_v\partial_i A_u-2\partial_v(\partial_iA_vh(\vec{x})\delta(u))=0.
\end{equation}
In the region $u<\epsilon$ and $u>\epsilon$, the first two terms are finite. At $u=\epsilon$, we are searching for a solution where $A_v$ jumps and $A_u$ may contain a delta function $\delta(u)$. On the other hand, from the first equation of Eq.~\eqref{eq5}, $F_{uv}$ must be finite. Therefore the delta functions in $\partial_u A_v$ and $\partial_v A_u$ must cancel each other at $u=\epsilon$.  Applying this condition, Eq.~\eqref{eq7} can be written as
\begin{equation}
    -2\partial_u\partial_iA_v-2\partial_v(\partial_iA_vh(\vec{x})\delta(u))=0.
\end{equation}

Integrating over $u$ we obtain
\begin{equation}
    \partial_i A_v(v,\vec{x})|_{u=\epsilon+0^+}=\partial_i A_v(v-h(\vec{x}),\vec{x})|_{u=\epsilon-0^+}.
\end{equation}
Note that this shift in $v$ doesn't commute with derivatives in the transverse directions, so this simple shift rule only applies to $\vec{\nabla}A$ instead of $A$ itself. The first equation in Eq.~\eqref{eq5} gives the same relation between $F_{uv}$ and $A$ as in Eq.~\eqref{eq3}. Comparing with Eq.~\eqref{eq4}, we find that this simple shift rule also applies to $\vec{\nabla} \phi(v,\vec{x})$. Hence,
\begin{equation}\label{eq31}
    \vec{\nabla}\phi(v,\vec{x})|_{u=\epsilon+0^+}=\vec{\nabla}\phi(v-h(\vec{x}),\vec{x})|_{u=\epsilon-0^+},
\end{equation}
which is the reason to write the inner product in the form of Eq.~\eqref{eq6}. In summary, after scattering with the shockwave, the wave function sourced by $J_0^R$ is modified according to the rule Eq.~\eqref{eq31}. If we choose to evaluate the inner product, Eq.~\eqref{eq6}, just after the scattering, along $u=\epsilon+0^+$, then $\phi_1$ and $\phi_2$ (in Eq.~\eqref{eq6}) should be
\begin{equation}
\begin{split}
   \vec{\nabla}\phi_1(v,\vec{x})&=\vec{\nabla}\phi^L(v,\vec{x}),\\
   \vec{\nabla}\phi_2(v,\vec{x})&=\vec{\nabla}\phi^R(v-h(\vec{x}-\vec{x_1}),\vec{x}).
\end{split}
\end{equation}

\subsection{Calculation of OTOC}
In this section, we evaluate the inner product in Eq.~\eqref{eq6}. It turns out that the relevant integral in is easier to do in momentum space. Define
\begin{equation}
    \phi(p,\vec{x})= \int dv e^{i p v}\phi(v,\vec{x}),
\end{equation}
so that
\begin{equation}
\begin{split}
\vec{\nabla} {\phi_1}(p,\vec{x})=&\int \frac{d^d \vec{k}}{(2\pi)^d} \frac{-i \vec{k}}{2\sin(\frac{\beta Dk^2}{2})} \frac{ p^{\frac{\beta Dk^2}{2\pi}-1}e^{Dk^2 t_2}}{\Gamma(\frac{\beta}{2\pi}Dk^2)} e^{ipe^{\frac{2\pi}{\beta}t_2}+ip h(\vec{x}-\vec{x}_1)}e^{i\vec{k}\cdot (\vec{x}-\vec{x_2})}\theta(p),   \\
\vec{\nabla} {\phi_2}(p,\vec{x})=&\int \frac{d^d \vec{k}}{(2\pi)^d} \frac{-i \vec{k}}{2\sin(\frac{\beta Dk^2}{2})} \frac{ p^{\frac{\beta Dk^2}{2\pi}-1}e^{Dk^2 t_2}}{\Gamma(\frac{\beta}{2\pi}Dk^2)} e^{-ipe^{\frac{2\pi}{\beta}t_2}}e^{i\vec{k}\cdot (\vec{x}-\vec{x_2})}\theta(p).
\end{split}
\end{equation}
Note that if we only keep the leading order terms in $k^2$, then the term $\sin(\frac{\beta Dk^2}{2})$ will cancel with $\Gamma(\frac{\beta}{2\pi}Dk^2)$. As long as we are considering large transverse coordinate separation, this approximation should be qualitatively correct. Plugging these into Eq.~\eqref{eq6} and approximating $\frac{R}{R_+}\sim 1$, we obtain
\begin{equation}\label{eq35}
\begin{split}
  & (A_1,A_2)=\\
   & \frac{1}{\pi}\int dp \int d^d\vec{x} \int \frac{d^dk}{(2\pi)^d} \int \frac{d^dk'}{(2\pi)^d} (\vec{k} \cdot \vec{k'}) p p^{Dk^2-1}e^{Dk^2 t_2} p^{Dk'^2-1}e^{Dk'^2 t_2}e^{-2ipe^{t_2}}e^{-iph(\vec{x}-\vec{x_1})}e^{i(\vec{k'}-\vec{k})\cdot (\vec{x}-\vec{x_2})}.
\end{split}
\end{equation}
To save space, we have suppressed the factor $\frac{\beta}{2\pi}$ setting the units of time. After some changes of variable and approximations shown in Appendix~\ref{detailB}, we obtain the final expression for the OTOC as
\begin{equation}\label{photonotoc}
\begin{split}
    (A_1,A_2) \sim \frac{1}{\left[\ln{\left(2+\frac{\Delta_O}{N}e^{\frac{2\pi}{\beta}t_{12}-\mu|\vec{x}_{12}|}\right)}\right]^{\frac{d}{2}}}.
\end{split}	
\end{equation}

At early time, this expression admits a large $N$ expansion in which the leading term still grows exponentially with time. Moreover, one can identify the same Lyapunov exponent and butterfly velocity as in the non-conserved OTOC in Eq.~\eqref{ineq2}. In the late time limit, the $\ln$ of the exponentially growing part in the denominator gives rise to a power law time decay behavior. Hence, we find a significant difference from the OTOC of non-conserved operators in the late time regime.

\section{Physical interpretation}\label{sec3}
In this section we try to understand the result in Eq.~\eqref{photonotoc} in a more intuitive way. In the last section, we calculated the OTOC by integrating first over the radial momentum. Although this makes the calculation easier, the physical reason why we expect a power law decay at late time is somewhat obscured. As an alternative approach, we can perform the spatial momentum integral at the beginning and rewrite the integral in Eq.~\eqref{eq35} as
\begin{equation}\label{physicaleq}
    \int_0^{+\infty} ds \int d^d\vec{x} \left|\vec{\nabla}\left(\frac{1}{(Ds)^{\frac{d}{2}}}e^{\frac{-|\vec{x}|^2}{4Ds}}\right)\right|^2 e^{-2ie^{-s}}e^{-i\frac{1}{N}e^{t_{12}-s-\mu|\vec{x}-\vec{x}_{12}|}}
\end{equation}
Here $s$ is defined as $s=\ln(\frac{e^{-t_2}}{p})$ and we have cut-off the large momentum contribution for $p>e^{-t_2}$.

This formula has a direct physical interpretation. The squared term can be viewed as the photon's wave function. From the solution Eq.~\eqref{eq10}, we see that the photon's wave function is extended both in the radial direction and in the transverse directions. Although we obtain Eq.~\eqref{physicaleq} in momentum space, it's more inspiring to think of $s$ as the radial coordinate. Small values of $s$ correspond to regions close to the horizon, while larger values of $s$ correspond to regions further from the horizon. The last term is a phase induced by the gravitational scattering, and $e^{t_{12}-s}$ measures the relative scattering energy of the two particles. Therefore, it is natural to think of the scattering between the photon's large wave-packet and the scalar particle's localized wave-packet as taking place over a large range of radial depths. Then as the scalar particle scans through the photon's wave function, the colliding energy effectively becomes smaller and smaller.

The term in the middle can be thought of as a regulator for large momentum (small $s$). Since we didn't obtain the complete solution of the photon's wave function in the black hole geometry, the regulator might be replaced by a more general function in the full solution. However, if we're only interested in the late time behavior, meaning $t_{12}-\mu|\vec{x}_{12}| \gg \ln(N)$, then the integral in $s$ receives its dominate contribution from $s> t_{12}$ due to the fast oscillation of the phase term for small $s$. The regulator is therefore not important, and we can directly see that the result is proportional to $\frac{1}{{t_{12}}^{\frac{d}{2}}}$.

\section{Correlator with stress-energy tensor}\label{sec4}

The low energy hydrodynamics of the stress-energy tensor shares some similarity with that of $U(1)$ charge. In this section, we show that the OTOC has the same late time behavior. As shown in \cite{hydrodynamics}\cite{sound}\cite{Son:2007vk}, in hydrodynamic limit, small perturbation of stress-energy tensor splits into sound and shear modes. The corresponding graviton wave functions in the bulk have a sound pole and a diffusive pole, respectively.  The shear mode is relatively easier to analyze, as it involves less components of the metric perturbation, but we will see that they have the same qualitative effect on the OTOC in the late time regime.

As above, we take the $\vec{k}$ direction as the $x$-axis. The shear modes consist of $T_{ty}$, $T_{xy}$. Here $y$ can be any direction perpendicular to $x$,  Sound modes are more involved, including $T_{tt}$, $T_{tx}$, $T_{xx}$ and $T_{yy}$. The shear and sound mode operators on the boundary excite bulk metric fluctuation corresponding to vector and scalar perturbations, respectively.  Taking the operator $T_{ty}$ as an example, it sources the bulk metric perturbations $h_{ty}$ and $h_{yr}$. Other choices of non-zero metric components are related to this by a gauge transformation. The spherically symmetric choice of scalar perturbation sourced by $T_{tt}$ involves $h_{tt}$, $h_{tr}$, $h_{xx}=h_{yy}$, and $h_{rr}$. For simplicity, we mainly discuss the shear mode and comment on the sound mode in the end.

\subsection{Wave function of graviton}

For a given momentum $\vec{k}$, choose the coordinate system such that the $x$-axis is parallel to $\vec{k}$. The shear modes involve the $T^{ty}(\omega,\vec{k})$ and $T^{xy}(\omega,\vec{k})$ components, and are described by
\begin{equation}
\begin{split}
     &T^{xy}=-D_T\partial_x T^{ty},\\
    &\partial_t T^{ty}+\partial_x T^{xy}=0,
\end{split}
\end{equation}
which together imply
\begin{equation}
    \partial_t T^{ty}=D_T\partial_x^2 T^{ty}.
\end{equation}

In AdS/CFT, the dynamics of these modes can be found by solving the linearized Einstein equation,
\begin{equation}
    \delta R_{\mu\nu}=\frac{2}{d}\Lambda h_{\mu\nu},
\end{equation}
where $d$ is the spacial dimension of boundary theory. $h_{\mu\nu}$ is the metric perturbation, $\delta g_{\mu\nu}=h_{\mu\nu}$. The equations are simplified if we set to zero all the components except $h_{ty}$ and $h_{ry}$. Then there are only two independent equations,
\begin{equation}
\begin{split}
    (k^2-\frac{\omega^2}{f(R)})h^y_r+\frac{i\omega}{f(R)}\partial_rh^y_t=0,\\
    \frac{d}{R}h^y_r+\frac{1}{f(R)}\partial_rh^y_r+\frac{i\omega}{f(R)}h^y_t=0,
\end{split}
\end{equation}
where $R$ is the radial coordinate in the metric Eq.~\eqref{eq0} and $r$ is the tortoise coordinate defined in Eq.~\eqref{eq01}. The two first order equations then lead to a second order differential equation for $h^y_t$,
\begin{equation}
    \partial^2_r h^y_t+\partial_r \ln\left(\frac{1}{R^d(\omega^2-k^2f)}\right)\partial_rh^y_t+(\omega^2-k^2f)h^y_t=0.
\end{equation}

This equation is the same as equation in $U(1)$ charge case except with $d-2$ replaced by $d$ (see Eq.~\eqref{eqa1}). Thus we have the same solution, except the diffusion constant $D_T=\frac{1}{d+1}R_+$ is different. Also, while the photon wave function excited by $J_0$ is spherically symmetric in the boundary spatial plane, in this case, since the shear mode operator $T_{ty}$ contains a spatial index, it breaks the spherical symmetry. So we expect that the wave function given by Eq.\eqref{eq10} only captures the dependence on directions $\vec{x}$ satisfying $\vec{x}\bot y$. In the $y$ direction, the mode propagates as sound. To avoid the complexity of mixing sound and shear modes, we restrict to consider the string operator living in $d-1$ spatial dimensions, $T^s(t,\vec{x}):=\int_{-\infty}^{+\infty} dy T_{ty}(t,y,\vec{x})$. Note that the choice of line operators over point operators can change late-time exponents by shifting the effective dimension of space.

\subsection{Interaction with the shockwave}

We continue to focus on the OTOC between energy-momentum tensor and a scalar operator with large conformal dimension. As above, we approximate the shockwave as sourced by the heavy scalar without backreaction from the graviton. Then it remains to consider the evolution of the graviton wave function in the geometry. In the following, to distinguish the metric perturbation from the shockwave, we will use $f(\vec{x})$ as the displacement in the metric,
\begin{equation}
    ds^2=2g_{uv}du[dv-\delta(u)f(\vec{x})du]+g_{xx}d\Omega_{d}.
\end{equation}
We solve the linearized equation
\begin{equation}
    \frac{1}{2}(D_{\rho}D_{\mu}h^{\rho}_{\nu}+D_{\rho}D_{\nu}h^{\rho}_{\mu}-D_{\rho}D^{\rho}h_{\mu\nu}-D_{\mu}D_{\nu}h^{\rho}_{\rho})-\frac{2}{d}\Lambda h_{\mu\nu}=0.
\end{equation}

This is a very complicated equation in general. Some other components have to be generated even if we start with only the shear mode perturbation. However, if we restrict the shockwave term $g_{uu}$ to only depend on directions that are perpendicular to $y$, the equation becomes easier to deal with. So we require $S$ to be a function of $x_i$ with $\vec{x}_i\bot y$.  Then the equations are simplified to
\begin{equation}
\begin{split}
    \partial_v[g^{uv}(\partial_vh_u^y-\partial_u h^y_v)]+\text{finite terms}=0,\\
    g^{uv}(\partial_v\partial_i h^y_u+\partial_u\partial_ih_v^y)+2g^{uv}\partial_v[\partial_i h^y_v f(\vec{x})\delta(u)]+\text{finite terms}=0,
\end{split}
\end{equation}
where `finite terms' denotes terms that don't contain a delta function. Here we are searching for a solution such that $\partial_u h^y_v$ and $h^y_u$ are proportional to $\delta(u)$. Requiring the cancellation of all $\delta(u)$s, we find that
\begin{equation}
   2 g^{uv}\partial_u\partial_i h^y_v+2g^{uv}\partial_v[\partial_i h^y_v f(\vec{x})\delta(u)]=0.
\end{equation}

Hence, we again have the simple shift rule,
\begin{equation}\label{greq2}
    \partial_ih^y_v(v,\vec{x})\rightarrow \partial_ih^y_v(v-f(\vec{x}),\vec{x}),
\end{equation}
after the graviton passing the shockwave. Finally, in order for $f$ to only depend on $\vec{x}\bot y$, we have to also consider a string operator built from the scalar $O^s(t,\vec{x})=\int dy O(t,\vec{x},y)$.

\subsection{Inner product}
As before, we construct the gauge invariant inner product using the symplectic form,
\begin{equation}
    \delta \mathcal{L}=\delta\phi \frac{\delta \mathcal L}{\delta \phi}+d\theta(\phi,\delta \phi).
\end{equation}
We find that
\begin{equation}\label{gravinner}
\begin{split}
   & (h_1,h_2)=\\
    &\int \sqrt{g} n^{\rho}[-h_1^{*\mu\nu}D_{\mu}{h_2}_{\nu\rho}+\frac{1}{2}h_1^{*\mu\nu}D_{\rho}{h_2}_{\mu\nu}+\frac{1}{2}{h_1^{*}}^{\nu}_{\rho}D_{\nu}{h_2}+\frac{1}{2}h_1^*D_{\nu}{h_2}^{\nu}_{\rho}-\frac{1}{2}h_1^*D_{\rho}h_2]-(1\longleftrightarrow 2).
\end{split}
\end{equation}

Knowing that the trace of $h_1$ and $h_2$ are zero, we only keep the first two terms. Then, after some cancellations, a much simpler form remains
\begin{equation}\label{greq3}
    (h_1,h_2)=\int \sqrt{\gamma} g^{xx}g^{uv}({h_1^{*}}_v^y\partial_{[v}{h_2}_{u]}^y-\Gamma^y_{yv}{h_1^{*}}_{[v}^y{h_2}_{u]}^y)-(1\longleftrightarrow 2).
\end{equation}
We have chosen $n$ to be $\frac{\partial}{\partial_v}$. This expression is very similar to the photon case (see Eq.~\eqref{eq12}), except for second term, which is proportional to the connection. Then we observe that $\Gamma^y_{yv}$ is of order $u$ near the horizon, so if we choose to evaluate the inner product along the hyper-surface $u\sim 0$, the second term can be neglected.

\subsection{OTOC of stress-energy tensor }
\subsubsection{Shear mode}
Just like the $U(1)$ case, the OTOC,
\begin{equation}
  Tr(\rho^{\frac{1}{2}}T^s_{ty}(t_2,\vec{x}_2)O^s(t_1,\vec{x}_1)\rho^{\frac{1}{2}}T^s_{ty}(t_2,\vec{x}_2)O^s(t_1,\vec{x}_1)),
\end{equation}
can be written as an inner product between the graviton wave functions, before and after passing the shockwave. Plugging the solution Eq.~\eqref{eq10} and Eq.~\eqref{greq2} into Eq.~\eqref{greq3}, we get exactly the same expression as in the $U(1)$ case, and the subsequent calculation is completely parallel. Note that these string operators live in an effective $d-1$ spacial dimension. The exponent of power law tail is also modified to $\frac{d-1}{2}$.

\subsubsection{Sound mode}
We can also consider a local insertion of a stress-tensor operator that creates spherically symmetric wave function propagating in the bulk.  At low energy, these modes have the dispersion relation of a sound wave. For instance, we can choose the insertion $T_{tt}$ and $\sum_{i=1}^d T_{ii}$. The pole in the Green's function is at
 \begin{equation}
     \omega=\pm v_s k-i\frac{d-1}{d} D_T k^2,
 \end{equation}
where $D_T$ is the diffusion constant (same as that of the shear mode). The quadratic term has the same effect as the shear mode, broadening the wave-packet. Hence, we expect the OTOC to have the same power law tail as in the photon case. Another interesting regime for the sound mode is at early time, where we have the shockwave that propagates at the butterfly velocity $v_B$ as well as the hydrodynamical mode that propagates at the sound speed $v_s$. For the gravity model we considered in this work, $v_B>v_s$ for physically sensible spatial dimension. However, in other models the sound speed might be larger.\footnote{For example, in \cite{PhysRevLett.117.091602,Blake_2016} there is a case where $v_B$ is small at low temperature.} In this case we find that the information carried by sound mode can scramble faster, the spreading speed of which is determined by $v_s$ instead.  We solve the sound mode gravitaional perturbation equation and provide a detailed analysis of the OTOC in Appendix~\ref{soundm}.

\section{Discussion and generalization }\label{sec5}
\subsection{Higher order corrections}
We have calculated the photon wave function by just keeping the leading order in $\omega$ and $k$. This gave an OTOC with a power law dependence on $t$ at large time. We may include higher order terms in the wave function, for example, consider
\begin{equation}
    \phi(\omega,k)= \frac{i(1+A \omega+B k)}{\omega+iDk^2+C \omega^2+D k^3}.
\end{equation}
The pole is now at
\begin{equation}
    \omega=-iDk^2(1+\gamma k).
\end{equation}
After Fourier transformation with respect to $\omega$, the wave function becomes
 \begin{equation}
 \begin{split}
     \phi(v,k)=&\int d\omega \phi(\omega,k)e^{-i\omega v},\\
     =&(1+Bk-iADk^2)v^{-Dk^2(1+\gamma k)}.
\end{split}
 \end{equation}

Following the steps above, we should multiply the integrand of Eq.~\eqref{eq35} by a factor $(1+\alpha k+\beta k^2)$. Also we should modify the exponent of $[2+f(\vec{x})e^{t_{12}}]$ in second equation of Eq.~\eqref{eq8} to $-Dk^2(1+\gamma k)$.   After performing integration with respect to $k$, these modifications only contribute factors of higher order in $t_{12}^{-1}$, and do not change the leading long time behavior. One can also include loop corrections due to a graviton, as we discuss in Appendix~\ref{nonlinear}. This correction creates a branch cut in the wave function but doesn't modify the long time behavior of the OTOC.

\subsection{Hydrodynamics and OTOCs of non-conserved operator}

\begin{figure}[H]
\centering
\begin{tikzpicture}
\begin{feynman}
\vertex (a) ;
\vertex [right=of a](b);
\vertex [below left=of a] (c);
\vertex [below left=of c] (u);
\vertex [above left=of a] (d);
\vertex [above right=of b] (e);
\vertex [below right=of b] (f);
\vertex [above right=of e] (j);
\vertex [below right=of f] (v);
\vertex [above=0.8cm of f] (p);
\vertex [above=0.1cm of e] (h){\(g_c\)};
\vertex [below=0.1cm of f] (g){\(g_c\)};
\vertex [left=0.1cm of a] (k){\(g_N\)};
\vertex [right=0.1cm of b] (l){\(g_N\)};
\vertex [left=0.1cm of c] (k){\(g_N\)};
\vertex [right=0.1cm of p] (s){\(g_N\)};
\diagram* {
(u)--(c)--(a)--(d),
(a)--[gluon](b),
(e)--[photon](b)--[photon](f),
(v)--(f)--(p)--(e)--(j),
(c)--[gluon](p)
    };
\end{feynman}
\end{tikzpicture}
\caption{intermediate hydrodynamic state in scalar-scalar scattering}
\label{scalarphoton}
\end{figure}
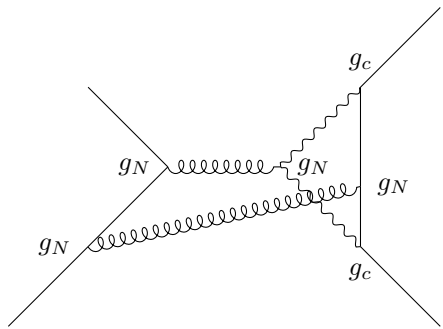

In this section, we explore the question of how hydrodynamic modes may affect scrambling of non-conserved operators. In the process shown in Fig~\ref{scalarphoton} (with the photon line representing either a photon mode or a sound mode), imagine that a massive scalar particle (created by a non-conserved operator) emits a hydrodynamic mode through a coupling of order $O(g_c)$ in the near boundary region. This mode grows in size as it falls into the black hole. The $in$ state before the shockwave scattering would contain a term $g_s\int f_s(p,x)f_m(p',x)|p,x'\rangle_s|p',x'\rangle_m|q,y\rangle$, where $f_s$ is the sound mode wave function and $f_m$ is the massive scalar wave function. The state with momentum $q$ is the second massive mode. The collidision induces a phase factor $e^{iqph(x,y)+iqp'h(x',y)}$. So we expect the scattering amplitude to be the one that involves hydrodynamic modes (like in Eq.~\eqref{eqhydrootoc}) multiplied with the one that involves only massive scalars. As discussed in the previous sections, the hydrodynamic OTOC has a power law tail at late time. However, the non-conserved OTOC is multiplied with it and the combination decays faster. Hence, the complete OTOC at late time is still controlled by massive particle scattering. On the other hand, the early time behavior can be modified, if we consider the non-conserved mode coupling with a sound mode, becuase the scattering amplitude between the sound mode and the second massive particle starts to decay earlier if $v_s>v_B$ (see Appendix~\ref{soundm} for details). Since the hydrodynamic mode's wave function is created with amplitude $O(g_c)$, its influence on the OTOC is of order $O(g_c^2)$. So even if $v_s>v_B$, the fast propagating wave-front cannot grow to exceed the same order,
\begin{equation}
OTOC\sim 1-\frac{c_1}{N}e^{t-\frac{|x|}{v_B}}-\frac{c_2g_c^2}{N}e^{t-\frac{|x|}{v_s}}.
\end{equation}

Since all modes couple to gravity, they also couple to the sound mode of the gravitational perturbation. In this case, $g_c^2\sim \frac{1}{N}$, and the last term can grow with time up to order $O(\frac{1}{N})$. A similar situation has been discussed in \cite{Lucas:2017ibu}, where they gave a bound on the squared commutator of the form
\begin{equation}
\text{squared commutator} \leq \frac{e^{t-\frac{|x|}{v_B}}}{N}+\frac{a(x,t)}{N}+O(\frac{1}{N^2}),
\end{equation}
where the function $a(x,t)$ is non-zero when $|x|<v_st$, and is bounded by an $O(1)$ quantity. However, the picture we considered here (Fig~\ref{scalarphoton}) is slightly different.

\subsection{Summary}
In conclusion, we have explored OTOCs between hydrodynamic operators and generic operators. In the late time regime, these OTOCs obey a power law scaling, while at early time the deviation still grows exponentially. In some models where the sound speed $v_s$ is large, the information near the wave-front scrambles with a velocity that depends on $v_s$. Finally, when generalizing to OTOCs of generic operators that couple to hydrodynamic modes, we found that the late time power law decay is absent, while there can be a small amount of information scrambling faster than $v_B$ (when $v_s>v_B$) near the wave-front.

We can also understand the conclusion intuitively in the boundary picture. Starting with a non-conserved operator $O$. A part of it evolves into $g_cJO$, where $J$ represents a hydrodynamic operator. Then a small part of the information is carried by the hydrodynamic mode. A pure hydrodynamic operator $J$ may spread fast (in sound mode case) but release its information slowly (which causes the late time power law tail). Therefore, although the order $g_c^2$ amount of information may propagate fast and lead to a rapidly moving wave-front in the OTOC, the scalar operator accompanied with $J$ releases most of the information and breaks the power law tail at late time. Due to conservation law constraints, the dynamics prevents $O$ from turning into pure $J$'s. In bulk theory, this is a constraint from gauge symmetry.

\section{Acknowledgements}

This work is supported in part by the Simons Foundation through the It From Qubit Collaboration.

\appendix

\section{Solution to the Maxwell-Einstein differential equation}

In our metric, $-g^{tt}=g^{rr}=\frac{R^2}{f}$ and $\sqrt{-g}=R^{-(d+2)}f$. The Maxwell equation simplifies to
\begin{equation}\label{eqa1}
    \partial^2_rA_t+\partial_r \ln \left(\frac{1}{R^{d-2}(\omega^2-k^2f)}\right)\partial_r A_t+(\omega^2-k^2 f)A_t=0
\end{equation}
There is a singular point at $f\rightarrow 0$, where the equation becomes
\begin{equation}
    \partial_r^2 A_t+\omega^2 A_t=0
\end{equation}
with solution $A_t(r,\omega,k)\sim e^{-i\omega r}$. We have picked the in-falling mode on this boundary. The ansatz
\begin{equation}
    A_t(r,\omega, k)=e^{-i\omega r}F(r,\omega,k)
\end{equation}
yields an equation for $F$,
\begin{equation}
    F''-2i\lambda\hat{\omega} F'-\partial_r \ln[R^{d-2}(\hat{\omega}^2-\hat{k}^2f)](F'-i\lambda\hat{\omega} F)+\lambda^2(\hat{\omega}^2-\hat{k}^2f)F=0,
\end{equation}
where $(\omega,k)=(\lambda \hat{\omega},\lambda \hat{k})$ and $\lambda \ll 1$.

The solution $F$ should have an expansion in the form $F=F_0+\lambda F_1+\cdots$. The leading $F_0$ satisfies
\begin{equation}
\begin{split}
    F_0''-\partial_r \ln[R^{d-2}(\hat{\omega}^2-\hat{k}^2f)]F_0'=0,\\
    \implies F_0=C_0+C_1\int_0^r dr' R'^{d-2}(\hat{\omega}^2-\hat{k}^2f).
\end{split}
\end{equation}
$F_0$ goes like $C_1\hat{\omega}^2 r$ as $r\rightarrow -\infty$, so a regular solution should have $C_1=0$. Similarly, the first order term $F_1$ should satisfy
\begin{equation}
    F_1''-2i\hat{\omega}F_0'-\partial_r \ln[R^{d-2}(\hat{\omega}^2-\hat{k}^2f)](F_1'-i\hat{\omega}F_0)=0.
\end{equation}
The integration constant can be fixed by requiring regularity at horizon, then one obtains
\begin{equation}
    F_1=i\hat{\omega}C_0 \int_{r_0}^r dr' [1-R'^{d-2}(1-\frac{\hat{k}^2}{\hat{\omega}^2})f].
\end{equation}
To first order in $\lambda$, we find
\begin{equation}
    A_t(r,\omega,k)=C_0\{1+i\omega\int_{r_0}^r dr' [1-R'^{d-2}(1-\frac{\hat{k}^2}{\hat{\omega}^2})f]\}.
\end{equation}

\section{Details of the OTOC calculation}\label{detailB}
Start from the inner product,

\begin{equation}
\begin{split}
&(A_1,A_2)\\
& =   \int dp \int d^d\vec{x} \int \frac{d^dk}{(2\pi)^d} \int \frac{d^dk'}{(2\pi)^d} (\vec{k} \cdot \vec{k'}) p p^{Dk^2-1} p^{Dk'^2-1}e^{-ip\left[2+\frac{\Delta_O}{N}e^{t_{12}-\mu|\vec{x}-\vec{x}_{12}|}\right]}e^{i(\vec{k'}-\vec{k})\cdot \vec{x}} \\
&=  \int d^d\vec{x} \int \frac{d^dk}{(2\pi)^d}\int \frac{d^dk'}{(2\pi)^d} (\vec{k} \cdot \vec{k'}) \Gamma[D(k^2+k'^2)] \left[2+\frac{\Delta_O}{N}e^{t_{12}-\mu|\vec{x}-\vec{x}_{12}|}\right]^{-D(k^2+k'^2)}e^{i(\vec{k'}-\vec{k})\cdot \vec{x}}. \\
\end{split}
\end{equation}
As in the main text, we will just keep the leading order dependence in $k$ for the gamma function. We change the integration variables to $K=\frac{k+k'}{2}$ and $\kappa=\frac{k-k'}{2}$. This gives
\begin{equation}\label{eq8}
    \begin{split}
         (A_1,A_2)\sim &\int d^d\vec{x} \int \frac{d^dK}{(2\pi)^d} \int \frac{d^d\kappa}{(2\pi)^d} 2^d\frac{K^2-\kappa^2}{2D(K^2+\kappa^2)} \left[2+\frac{\Delta_O}{N}e^{t_{12}-\mu|\vec{x}-\vec{x}_{12}|}\right]^{-2D(K^2+\kappa^2)}e^{-2i\kappa \cdot \vec{x}}\\
          = &\int \frac{d^dx}{(4\pi)^d}\left[\frac{d}{D^{d+1}}E(\ln{g},\frac{|\vec{x}|^2}{2D})-\frac{1}{2D^{d+1}}\frac{1}{(\ln{g})^d}e^{-\frac{|\vec{x}|^2}{2D \ln g}}\right],
    \end{split}
\end{equation}
where $g$ and $E$ are defined as
\begin{equation}
\begin{split}
g=2+\frac{\Delta_O }{N}e^{t_{12}-\mu|\vec{x}-\vec{x}_{12}|},\\
E(z,a)=\int_{z}^{\infty} dy \frac{1}{y^{d+1}}e^{-\frac{a}{y}}.
\end{split}
\end{equation}

Since both of the two terms contain $e^{-\frac{|\vec{x}|^2}{2D \ln g}}$, we expect the integral to receive its dominant contribution from $|x|\sim 0$. Integrating $\vec{x}$ over this saddle point gives a factor of $(2D \ln g)^{\frac{d}{2}}$. Therefore, we finally obtain
\begin{equation}\label{Bresult}
(A_1,A_2)\propto \frac{1}{[\ln{(2+\frac{\Delta_\mathcal{O}}{N}e^{\frac{2\pi}{\beta}t-\mu|x_{12}|})}]^{\frac{d}{2}}}.
\end{equation}

In fact, the saddle point approximation in last step is only valid for very small diffusion constant $D$ and for not too large values of the function $g(x)$. Thus it is necessary to discuss the late and early time limits separately from the above treatment. Looking back at Eq.~\eqref{eq8}, in the large $t_{12}$ limit, $t_{12}\gg Ln N+\frac{|\vec{x}_{12}|}{v_B}$. We can expand the function $\ln{g}\sim t_{12}(1-\frac{|\vec{x}|}{v_Bt_{12}})$, for $|x|< cv_B t$, where $c$ is some finite constant smaller than $1$. Then the integral over $x$ can be evaluated as
\begin{equation}
\begin{split}
&\int d^d\vec{x} \frac{1}{(\ln g)^{\alpha}}e^{-\frac{|x|^2}{2D\ln{g}}},\\
=&\int^{|x|<cv_Bt} d^d\vec{x}\frac{1}{t_{12}^\alpha(1-\frac{|x|}{t_{12}})^{\alpha}}e^{-\frac{|x|^2}{2Dt_{12}}[1+O(\frac{|x|}{t_{12}}))]}+\int_{|x|>cv_Bt} d^d\vec{x} \frac{1}{(\ln g)^{\alpha}}e^{-\frac{|x|^2}{2D\ln{g}}}.
\end{split}
\end{equation}

For the first part, the change of variable to $x\rightarrow \frac{x}{\sqrt{2Dt_{12}}}$ gives
\begin{equation}
\int^{|x|<c\frac{v_B}{\sqrt{2D}}\sqrt{t}} d^d\vec{x}\ t_{12}^{\frac{d}{2}-\alpha}\ (1+O(\frac{|x|}{\sqrt{t_{12}}}))e^{-|x|^2(1+O(\frac{|x|}{\sqrt{t_{12}}}))}\sim t_{12}^{\frac{d}{2}-\alpha}(1+O(t_{12}^{-\frac{1}{2}})).
\end{equation}
The second part is bounded by
\begin{equation}
\int_{|x|>cv_Bt}d^d\vec{x} \frac{1}{(\ln{2})^{\alpha}}e^{-\frac{|x|^2}{2Dt_{12}}}\sim O(e^{-\frac{v_B^2t_{12}}{D}}).
\end{equation}
From these results, the late time amplitude indeed decays in a power law manner, and is consistent with the result Eq.~\eqref{Bresult}.
\begin{equation}
    (A_1,A_2)\sim\frac{1}{|t_{12}|^{\frac{d}{2}}}.
\end{equation}

On the other hand, when $t_{12}-\frac{|\vec{x}_{21}|}{v_B}\ll \ln{N}$, we expand $\ln{g}\sim \ln{2}+\frac{\Delta_{\mathcal{O}}}{2N}e^{t_{12}-\frac{|\vec{x}-\vec{x}_{12}|}{v_B}}$, as well as
\begin{equation}\label{eq9}
\begin{split}
    E(\ln{g},a)\sim & E(\ln{2},a)+\frac{\Delta_{\mathcal{O}}}{2N}e^{t_{12}-\frac{|\vec{x}-\vec{x}_{12}|}{v_B}}\frac{d}{d z}E(z,a)|_{z=\ln{2}},\\
    \sim & E(\ln{2},a)-\frac{1}{2N}e^{t_{12}-\frac{|\vec{x}-\vec{x}_{12}|}{v_B}}\frac{1}{(\ln{2} )^{d+1}}e^{-\frac{a}{\ln{2}}}.
\end{split}
\end{equation}
The leading order deviation is
\begin{equation}\label{Bearlytime}
\begin{split}
    (A_1,A_2)\sim \frac{1}{(\ln{2})^{\frac{d}{2}}}\left[1-\frac{1}{2N}\frac{\frac{d}{2}}{\ln {2}}e^{t_{12}}\int d^d\tilde{x} \frac{2}{3}\left(1+\frac{|x|^2}{d}\right)e^{-\frac{|\tilde{x}-{\tilde{x}_{12}/\sqrt{D}}|}{v_B/\sqrt{D}}-|\tilde{x}|^2   }\right]
\end{split}
\end{equation}
This matches with the result Eq.~\eqref{Bresult}, when expanding around small values of $D$, because the integral in Eq.~\eqref{Bearlytime} is approximated by $e^{-\frac{|x_{12}|}{v_B}}+e^{-\frac{|x_{12}|^2}{D}}$. We can see that the velocity of the wave-front is still given by $v_B$. For completeness, we give the exact result of this integral in 1d,
\begin{equation}
e^{-\frac{|x_{12}|}{v_B}+\frac{D}{4v_B^2}}\text{erf}\left(\frac{\sqrt{D}}{2v_B}-\frac{|x_{12}|}{\sqrt{D}}\right)\left(\frac{1}{2}+\frac{D}{12v_B^2}\right)-\left(\frac{\sqrt{D}}{6v_B}+\frac{1}{3}\frac{|x_{12}|}{\sqrt{D}}\right)e^{-\frac{|x_{12}|^2}{D}}.
\end{equation}

\section{OTOCs of sound mode operators}\label{soundm}
In this section, we evaluate the OTOC between a sound mode operator (eg. $T_{tt}$, $T_{xx}+T_{yy}$) and a scalar operator with large conformal dimension $\Delta$. For this purpose, we first solve the sound mode equation in a symmetric gauge to second order of $\omega$ and $k$. By doing this, we obtain the dispersion relation as in \cite{Policastro_2002} and the near horizon form of the wave function. Unlike in \cite{Policastro_2002}, we apply a gauge fixing condition that preserves boundary spatial rotational symmetry.

If the boundary source respects rotational symmetry (for example, consider a boundary insertion of $T_{tt}$ or $\sum_{i=1}^d T_{ii}$), we expect a bulk configuration satisfying $h_{ti}=h_{ri}=0$ and $h_{ij}$ proportional to identity matrix, for $i,j\in \{1,\cdots, d\}$.  Without loss of generality, we will take $d=2$ in the following. Then the non-zero components are $h_{tt}$,$h_{tr}$, $h_{rr}$ and $h_{xx}=h_{yy}$. They satisfy a set of differential equations. Using the tortoise coordinate $r=-\int \frac{dR}{f(R)}$, and the Fourier decomposition $h_{MN}(r,t,x)=\int d\omega dk h_{MN}(r,\omega,k)e^{-i \omega t+ikx}$, these equations are written as

\begin{equation}\label{seq1}
-i (f+3) R \omega  h_{\text{xx}} +k^2 R^2 h_{\text{tr}} +2 i R^2 \omega  h_{\text{xx}}' -2 i R \omega  h_{\text{rr}} =0,
\end{equation}

\begin{equation}\label{seq2}
   -\frac{k R^2 \omega  h_{\text{rr}} }{f}+\frac{i k R^2 h_{\text{tr}}' }{f}-k R^2 \omega  h_{\text{xx}} =0,
\end{equation}

\begin{equation}\label{seq3}
   \frac{k R \left(h_{\text{rr}} -h_{\text{tt}} \right)}{f}=0,
\end{equation}

\begin{equation}\label{seq4}
\begin{split}
    \frac{2 f^2 h_{\text{xx}} }{R}+\frac{(f+3) h_{\text{rr}} }{2 R}-\frac{(f+3) h_{\text{tt}} }{2 R}-f h_{\text{xx}}' +i \omega  h_{\text{tr}} +h_{\text{tt}}' =0,
\end{split}
\end{equation}

\begin{equation}\label{seq5}
    \begin{split}
       -\frac{\left(f+k^2 R^2+3\right) h_{\text{rr}} }{R^2}+\left(\frac{3-f}{R^2}+k^2\right) h_{\text{tt}} +  \left(-f k^2-\frac{6 f}{R^2}+\omega ^2\right)h_{\text{xx}}-\frac{h_{\text{rr}}' }{R}-\frac{2 i \omega  h_{\text{tr}} }{R}-\frac{h_{\text{tt}}' }{R}+h_{\text{xx}}'' =0,
    \end{split}
\end{equation}

\begin{equation}\label{seq6}
\begin{split}
    h_{\text{tt}}''+\frac{(7 f-9) h_{\text{tt}}'}{2 R}+\frac{(f-3) f h_{\text{xx}}'}{R}-\frac{(f-3) h_{\text{rr}}'}{2 R}+2 i \omega  h_{\text{tr}}'+\cdots=0,
\end{split}
\end{equation}

\begin{equation}\label{seq7}
   h_{\text{tt}}''-2 f h_{\text{xx}}''+\frac{3 (f-3) h_{\text{tt}}'}{2 R}+\frac{3 f (f+1) h_{\text{xx}}'}{R}+\frac{3 (f+1) h_{\text{rr}}'}{2 R}+2 i \omega  h_{\text{tr}}'+\cdots=0.
\end{equation}

These seven equations are not independent. They reduce to three independent first order differential equations together with an algebraic equation, Eq.~\eqref{seq3}. Then the last three equations, Eq.~\eqref{seq5}-\eqref{seq7}, give an algebraic constraint,

\begin{equation}\label{seq8}
    \omega  h_{xx}(r) \left(3 f^2-2 f \left(k^2 R^2+6\right)+4 R^2 \omega ^2+9\right)+\omega h_{tt}(r) \left(-6 f+2 k^2 R^2+6\right)-i R h_{tr}(r) \left((f-3) k^2+4 \omega ^2\right)=0.
\end{equation}

There are two independent solutions, with out-going and in-falling conditions near the horizon. To see this, make the substitution $\lim_{r\rightarrow -\infty}h_{\mu\nu}(r)=e^{\nu r}F_{\mu\nu}$ to the equations and take the limit $R\rightarrow 1$. The equations become
\begin{equation}
\nu\left(
    \begin{array}{c}
    F_{xx}    \\
    F_{tr}     \\
    F_{tt}
    \end{array}
    \right)=
    \left(
    \begin{array}{ccc}
    \frac{3}{2} & \frac{ik^2}{2\omega}  & 1  \\
    0     & 0 & -i\omega \\
    0 & -i\omega & 0
    \end{array}
    \right) \left(
    \begin{array}{c}
    F_{xx}    \\
    F_{tr}     \\
    F_{tt}
    \end{array}
    \right)
\end{equation}
There are three eigenvalues, $(\frac{3}{2},i\omega,-i\omega)$, corresponding to a spurious solution, out-going, and in-falling solutions, respectively. The eigenvalue $\frac{3}{2}$ is discarded, since the corresponding eigenvector is not compatible with the constraint Eq.~\eqref{seq8}. The eigenvector of the in-falling solution tells us that
 \begin{equation}\label{seq9}
     \left(
     \begin{array}{c}
        F_{xx}    \\
        F_{tr}  \\
        F_{tt}
     \end{array}
     \right)=C(\omega,k)\left(
     \begin{array}{c}
        -\frac{k^2-2 i \omega }{\omega  (2 \omega -3 i)}  \\
          1 \\
          1
     \end{array}
     \right).
 \end{equation}
One can check that the gauge condition fixes the gauge completely. As a result, the in-falling solution is unique. To second order in momentum, we find the solution for $h_{xx}$
\begin{equation}
\begin{split}
    h_{xx}(r,\omega,k)=C(\omega,k)e^{-i \omega r}[-\frac{2}{3R}+\lambda\int^R_{c_x^1} dR'\frac{i(k^2 f-2\omega^2(1-R'^2))}{3R'^2f\omega}+\lambda^2 h_{xx}^{(2)}+O(\lambda^3)]\\
    h^{(2)}_{xx}=\frac{1}{R}\int^R_{c_x^2} dR'[\frac{k^2 f-2\omega^2}{9R'^2f}Ln(\frac{f}{3})+\frac{4(k^2-\omega^2)-k^2R'^3}{9R'^2f}]- \left(\frac{4 \omega ^2}{9}-\frac{k^2}{3}\right)r-\frac{ \omega ^2r^2}{3}
\end{split}
\end{equation}
The integration constants $c_x^1$ and $c_x^2$ are fixed by requiring that the $r\rightarrow -\infty$ limit of $h_{xx}$ matches with Eq.~\eqref{seq9}.

The solution for $h_{tt}$ is
\begin{equation}
\begin{split}
    h_{tt}(r.\omega,k)=C(\omega,k)e^{-i \omega r}[\frac{1}{3}R^2+\frac{2}{3R}+\lambda\frac{2+R^3}{R}\int^{R}_{c_{t}^1}dR'\frac{3i\omega R'^4}{(2+R'^3)^2f}+\lambda^2h^{(2)}_{tt}+O(\lambda^3)],
\end{split}
\end{equation}
where $h_{tt}^{(2)}$ is a complicated function. Again, the integration constants can be chosen according to Eq.~\eqref{seq9}.
Finally, solution for $h_{tr}$ is
\begin{equation}
    h_{tr}(\omega,k)=C(\omega,k)e^{-i\omega r}[1-i\lambda\omega\int^R_1 dR'\frac{1-R'^2}{f}+\lambda^2(-i\omega\int dR'\frac{h_{tt}^{(1)}+fh_{xx}^{(1)}}{f}+\frac{1}{2}\omega^2r^2)+O(\lambda^3)].
\end{equation}

Close to the boundary ($r\rightarrow 0$), $h_{tt}\sim O(\frac{1}{r})$, and $h_{xx}$ has the asymptotic form
\begin{equation}
    h_{xx}\rightarrow \frac{(1+\frac{i\omega}{3}Ln(3))(2\omega^2-k^2)+\frac{4i\omega}{3}(k^2-\omega^2)}{-3i\omega r^2}+O(\frac{1}{r}).
\end{equation}
Therefore, for the boundary souce $h^0_{xx}(\omega,k)=h^0_{yy}(\omega,k)=1$, we obtain the overall constant $C(\omega,k)=\frac{-3i\omega }{(1+\frac{i\omega}{3}Ln(3))(2\omega^2-k^2)+\frac{4i\omega}{3}(k^2-\omega^2)}$. The near horizon limit of all the components is also clear. Due to rotational symmetry, the $x$-axis can be any direction. So we have the following ansatz for the wave function, with the correct sound pole,
\begin{equation}
    h_{tt}(r,t,\vec{x})=h_{rr}(r,t,\vec{x})\sim\int d\omega d^2\vec{k}\frac{-\frac{3}{2}i\omega(\frac{1}{3}R^3+\frac{2}{3R}) }{(\omega^2-\frac{1}{2}k^2)+\frac{i\omega}{3}k^2}e^{-i\omega (t+r)+i\vec{k}\cdot\vec{x}},
\end{equation}
\begin{equation}
    h_{xx}(r,t,\vec{x})\sim \int d\omega d^2\vec{k}\frac{1}{R}\frac{i\omega }{(\omega^2-\frac{1}{2}k^2)+\frac{i\omega}{3}k^2}e^{-i\omega (t+r)+i\vec{k}\cdot\vec{x}},
\end{equation}
\begin{equation}
    h_{tr}(r,t,\vec{x})\sim\int d\omega d^2\vec{k}\frac{-\frac{3}{2}i\omega }{(\omega^2-\frac{1}{2}k^2)+\frac{i\omega}{3}k^2}e^{-i\omega (t+r)+i\vec{k}\cdot\vec{x}}.
\end{equation}

Using this, we can estimate the late time tail of the sound mode OTOC, given that the inner product contains a term
\begin{equation}
    \int \sqrt{\gamma} dv n^v ({h_1}_{xx}^*(v)\partial_v{h_2}_{xx}(v)-{h_2}_{xx}(v)\partial_v{h_1^*}_{xx}(v)),
\end{equation}
where ${h_2}_{xx}(v)={h_1}_{xx}(v-h)$ is a solution to the linearized Einstein equation on the shockwave background. The calculation is slightly different from the previous one due to the linear in $k$ term in the sound pole dispersion. In the end, it becomes the following integral
\begin{equation}
\begin{split}
&(h_1,h_2)\\
\propto &\int d^d\vec{x}\int \frac{d^d\vec{k}}{(2\pi)^d}\frac{d^d\vec{k'}}{(2\pi)^d}\int \frac{dp}{2\pi} pp^{Dk^2-1}\cos[v_s k \ln(p)]p^{Dk'^2-1}\cos[v_sk'Ln(p)]G(p)e^{-\frac{\Delta}{N}e^{t_{12}-\mu|\vec{x}-\vec{x}_{12}|}}e^{-i(\vec{k}-\vec{k'})\cdot \vec{x}},
\end{split}
\end{equation}
where we have introduced an unknown cutoff factor $G(p)$ to regulate the large momentum behavior. Similar to the method in Section~\ref{sec3}, one defines $s=-\ln(p)$ and performs the integral over the transverse momentum first. Then the inner product can be written as
\begin{equation}
\begin{split}
  &  \int d^d\vec{x}\int_0^{+\infty}ds \frac{1}{s^{d+1}}(e^{-\frac{(|\vec{x}|-v_s s)^2}{4Ds}}+e^{-\frac{(|\vec{x}|+v_s s)^2}{4Ds}})^2e^{-i\frac{\Delta}{N}e^{t_{12}-s-\frac{|\vec{x}-\vec{x}_{12}|}{v_B}}}G(s),\\
  \sim & \int d\Omega_d\int_0^{+\infty} ds \frac{1}{s^{\frac{d}{2}+1}}e^{-i\frac{\Delta}{N}e^{t_{12}-s-\frac{|v_s s \hat{n}-\vec{x}_{12}|}{v_B}}}G(s).
\end{split}
\end{equation}
In the final, we performed the integral over $|\vec{x}|$ assuming that it receives its dominant contribution around $|\vec{x}|\sim v_s s$. The factor $G(s)$ regulates small $s$ region.  When $t_{12}$ is large, the integral in $s$ becomes significant only when $s>t_{12}$. So we conclude that the late time OTOC has a power law tail.

It's also interesting to discuss the early time regime, where we can expand the integrand in $\frac{1}{N}$, and obtain
\begin{equation}
\text{OTOC}  \sim  \int d\Omega_d \int_0^{+\infty} ds \frac{1}{s^{\frac{d}{2}+1}}G(s)\left(1-\frac{i\Delta}{N}e^{t_{12}-s-\frac{|v_ss \hat{n}-\vec{x}_{12}|}{v_B}}+\cdots\right)
\end{equation}
We need to analyze the integral over $s$. Denote the angle between $\hat{n}$ and $\vec{x}_{12}$ by $\theta$. The exponential reaches its maximum at $s_*=\frac{x_{12}}{v_s}(\cos\theta-\sin\theta \tan\varphi)$, with $\varphi=\arcsin\frac{v_B}{v_s}$. So for  $v_B>v_s$, or $\cos \theta<\frac{v_B}{v_s}$, then $s_*=0$, and we conclude that the information still scrambles at the speed $v_B$. On the other hand, for $\cos \theta > \frac{v_B}{v_s}$, we expand $s=s_*+l$ and approximate the exponential factor as
\begin{equation}
e^{t_{12}-s-\frac{|v_ss\hat{n}-\vec{x}_{12}|}{v_B}}\approx e^{t_{12}-\frac{x_{12}}{v_s}(\cos\theta+\sin\theta\cot\varphi)}e^{-\frac{v_s\cos^3\varphi}{2x_{12}\sin \theta \sin \varphi}l^2}.
\end{equation}
After integrating over $l$, the $O(\frac{1}{N})$ term becomes \footnote{Since $s_*$ can be very large, this result is not sensitive to the regulator $G(s)$. However, the constant term does depend on it, which will change the $O(\frac{1}{N})$ term's prefactor after normalization. }
\begin{equation}
\sim \frac{1}{N}(\frac{x_{12}}{v_s})^{-\frac{d+1}{2}}e^{t_{12}-\frac{x_{12}}{v_s}(\cos\theta+\sin\theta\cot\varphi)}.
\end{equation}
From this expression, we learn that when $v_s>v_B$, a portion of the information, corresponding to solid angle $\theta_0$, can scramble at a speed $\frac{v_s}{\cos\theta_0+\sin\theta_0\cot \varphi}>v_B$, with $\cos\theta_0<\sin \varphi=\frac{v_B}{v_s}$. For $\theta_0$ sufficiently small, the information spreading speed is close to speed of sound.

\section{Non-linear corrections}\label{nonlinear}

\begin{figure}[H]
\centering
\begin{tikzpicture}
\begin{feynman}
\vertex (a) {$t$};
\vertex [right=of a] (b);
\vertex [right=of b] (c);
\vertex [right=of c] (d) {$t$};
\vertex [right=0.8cm of b] (r);
\vertex [below=0.8cm of r] {$F_{xr}$};
\vertex [above=0.8cm of r] {$h_{tx}$};
\diagram*{
(a)--[photon](b),
(b)--[photon,half right](c),
(c)--[gluon,half right](b),
(c)--[photon](d),
};
\end{feynman}
\end{tikzpicture}
\caption{gravitational dressing to photon's wave function}
\label{figloop}
\end{figure}
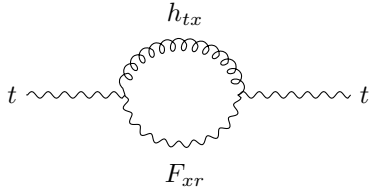

In this section, we consider the leading non-linear correction to the photon's propogator. The correction is from graviton dressing, as shown in Fig.~\ref{figloop}. Expanding the Einstein-Maxwell action, one obtains a gauge-gauge-graviton vertex of the following form,
\begin{equation}\label{eqAAg}
   \frac{1}{g_c^2} \int d^{d+2}x  A_{\mu} \partial_{\nu}[\sqrt{-g}(F^{\rho\nu}h^{\mu}_{\rho}+F^{\mu\rho}h^{\nu}_{\rho}-\frac{1}{2}F^{\mu\nu}h^{\rho}_{\rho})].
\end{equation}
Following \cite{CaronHuot:2009iq}, we consider the gravitational dressing in the near horizon region, and only include the interaction between the photon and the diffusive mode of the graviton. In other words, we only keep $h_{tx}$ and $h_{rx}$ non-zero. Many terms in Eq.~\eqref{eqAAg} are suppressed by small $\omega$ and $k$, so the leading order term involving $A_t$ is
\begin{equation}
\frac{1}{g_c^2}\int d^{d+2}x \sqrt{-g} g^{tt}g^{rr}\partial_rA_tF_{rx} h^{x}_{t}.
\end{equation}

To be consistent with the main text, we still use the gauge $A_x=0$. To find the bulk correlators, we have to solve the Maxwell-Einstein equation, Eq.~\eqref{eqmaxein}, with a source term such that
\begin{equation}
\begin{split}
    A_t=\int d^{d+2} \vec{x}\ G_{tt}(\vec{r}-\vec{r}')J^t(\vec{r}'),\\
    A_r=\int d^{d+2} \vec{x}\ G_{rr}(\vec{r}-\vec{r}')J^r(\vec{r}'),
\end{split}
\end{equation}\
where $J^t$ and $J^r$ are 4-vectors that satisfy
\begin{equation}
    \partial_t J^t+\partial_rJ^r+\partial_x J^x=0.
\end{equation}

To obtain $G_{tt}$, we require $J^r=0$, and $J^t(r,\omega,k)=\delta(r-r')$. Then $J^x(r,\omega,k)=\frac{\omega}{k}\delta(r-r')$. One can thenderive the following equation,
\begin{equation}\label{eqa1}
    \partial^2_rA_t+\partial_r \ln\left(\frac{1}{R^{d-2}(\omega^2-k^2f)}\right)\partial_r A_t+(\omega^2-k^2 f)A_t=\frac{(\omega^2-k^2f)R^{d-2}}{k^2}\delta(r-r').
\end{equation}
Applying the in-falling condition at the horizon and a Neumann boundary condition on the boundary, we obtain an approximate solution by expanding in small $\omega$ and $k$. Moreover, if we focus on the near horizon region, the solution takes a simple form,
\begin{equation}
\begin{split}
    &G_{tt}(\omega,k,r,r')\sim \frac{\frac{\omega^2}{k^2}}{i\omega-Dk^2}(1+O(\omega r)),\\
    &G_{tx,tx}(\omega,k,r,r')\sim \frac{\frac{\omega^2}{k^2}}{i\omega-D_Tk^2}(1+O(\omega r)).\\
\end{split}
\end{equation}

Similarly, to solve for $G_{rr}$, we use a source such that $J_t=0$, and $J_r(r,\omega,k)=\delta(r-r')$. So we have $J^x(r,\omega,k)=\frac{1}{-ik}\delta(r-r')$, as well as the equation
\begin{equation}
\partial^2_r\tilde{A}_r + \partial_rLog(R^{d-2})\partial_r \tilde{A}_r+(\omega^2-k^2f)\tilde{A}_r=\frac{\omega^2}{k^2}\delta(r-r'),
\end{equation}
where $\tilde{A}_r$ contains a contact term and is defined as $\frac{1}{R^{d-2}}A_r+\frac{1}{k^2}\delta(r-r')$. In the near horizon region, one can also find that
\begin{equation}
G_{rr}(r,r',\omega.k)\sim \frac{\frac{\omega^2}{k^2}}{i\omega-Dk^2}(1+O(\omega r)).
\end{equation}

To simplify the calculation, we replace $\omega$ by its value at the pole and set $\frac{\omega^2}{k^2}$ to $Dk^2$. Then we evaluate the bubble diagram by summing over imaginary frequencies and continuing to real frequency in the end,
\begin{equation}
\begin{split}
&\Sigma_{tt}(i\omega_n\rightarrow \omega+i0^+,k)=\frac{1}{N}\int \frac{d^dk'}{(2\pi)^d}k'^2T\sum_n\mathcal{G}_{rr}(i\omega'_n,k')\mathcal{G}_{xt,xt}(i\omega_n-i\omega_n',k-k')|_{\i\omega_n\rightarrow\omega+i0^+}\\
\approx &\frac{1}{N}\int \frac{d^dk'}{(2\pi)^d}\int \frac{d\omega'}{2\pi}k'^2[\frac{\Im{G_{rr}(\omega',k')}}{\omega'}G_{tx,tx}(\omega-\omega',k-k')+G_{rr}(\omega',k')\frac{\Im G_{tx,tx}(\omega-\omega',k-k')}{\omega-\omega'}]\\
\sim& \frac{c}{N} k^2\omega[-i\omega+D_m k^2]^{\frac{d}{2}-1},
\end{split}
\end{equation}
where $D_m=\frac{DD_T}{D+D_T}$ is a mixed diffusion constant. The loop correction creates a branch cut along $-i\omega<-D_m k^2$, as expected by considering the on-shell condition of the two internal states. Then we suppose that the wave function in the near horizon region also receives a correction in the denominator, of the form
\begin{equation}
A_t\sim \frac{\omega}{\omega+iDk^2-\frac{c}{N}k^4 \omega[-i\omega+D_m k^2]^{\frac{d}{2}-1}}.
\end{equation}

The branch cut structure is similar to that considered in \cite{Chen-Lin:2018kfl}. Close to the line $-i\omega<-D_m k^2$, the pole is split into two poles, at $\omega=-iDk^2\pm \frac{\alpha}{N} |k|^{d+4}$. So the modification to real time functions is sub-leading in $1/|t|$. When going to the real time, we need to add a integral along a contour that circulates the branch cut. This integral is proportional to
\begin{equation}
\begin{split}
&\int_{D_mk^2}^{+\infty}ds \frac{c\frac{k^4}{N}s(s-D_mk^2)^{\frac{d}{2}-1}e^{-st}}{(s- Dk^2)^2+c^2\frac{k^8s^2}{N^2}(s-D_mk^2)^{d-2}}.\\
\end{split}
\end{equation}
The integrand is like a Lorentzian function, so we can estimate it by multiplying its peak height and width, which is again $e^{-Dk^2 t}(1+O(k^2))$. So we conclude that the loop correction doesn't change the conclusion in the main text.

\bibliographystyle{unsrt}
\bibliography{reference}
\end{document}